\documentclass[conference,compsoc]{IEEEtran}
\ifCLASSOPTIONcompsoc
  \usepackage[nocompress]{cite}
\else
  \usepackage{cite}
\fi

\usepackage{times}
\usepackage{graphicx}
\usepackage{subfigure}
\graphicspath{{figs/}}
\usepackage{epstopdf}		
\usepackage[cmex10]{amsmath}
\usepackage{longtable}
\usepackage{pdfpages}
\usepackage{bbding}
\usepackage{pifont}
\usepackage{wasysym}
\usepackage{amsmath}
\usepackage{listings}
\usepackage{url}
\usepackage{color}
\usepackage{setspace}
\usepackage{float}
\usepackage{multirow}

\begin{document}
%
\title{
Generating Labeled Flow Data from MAWILab Traces for Network Intrusion Detection
}


\author{
Jinoh Kim\IEEEauthorrefmark{1}\, 
Caitlin Sim\IEEEauthorrefmark{2}\,
Jinhwan Choi\IEEEauthorrefmark{1}\ 
\IEEEauthorblockA{\\
Texas A\&M University, Commerce, TX 75428 \IEEEauthorrefmark{1}\\
Dougherty Valley High School, San Ramon, CA 94582 \IEEEauthorrefmark{2}\\
Email: jinoh.kim@tamuc.edu, caitlinsim@gmail.com, jchoi8@leomail.tamuc.edu}}

\maketitle

\begin{abstract}
A growing issue in the modern cyberspace world is the direct identification of malicious activity over network connections.
The boom of the machine learning industry in the past few years has led to the increasing usage of machine learning technologies, which are especially prevalent in the network intrusion detection research community.
When utilizing these fairly contemporary techniques, the community has realized that datasets are pivotal for identifying malicious packets and connections, particularly ones associated with information concerning labeling in order to construct learning models. 
However, there exists a shortage of publicly available, relevant datasets 
to researchers in the network intrusion detection community. 
Thus, in this paper, we introduce a method to construct labeled flow data by combining the packet meta-information with IDS logs to infer labels for intrusion detection research.
Specifically, we designed a NetFlow-compatible format due to the capability of a a large body of network devices, such as routers and switches, to export NetFlow records from raw traffic.
In doing so, the introduced method at hand would aid researchers to access relevant network flow datasets along with label information. 
\end{abstract}

\section{Introduction}

With the growing intensity of cyber attacks, it is becoming even more crucial to identify this malicious activity in a timely manner. 
One such technique is scanning individual packets for textual patterns (also known as "signatures") that have been previously collected from other attack packets~\cite{vasilomanolakis2015taxonomy}.
Although this idea is appealing due to its high detection accuracy, its limitations cannot be ignored, namely various encryption and privacy issues. 
Another approach in identifying potential network intrusions is the use of machine learning that relies on statistical information without analyzing internal payload~\cite{buczak2016survey}.
Due to the rapid, modern development of machine learning technologies, this alternative approach has been widely accepted by the network intrusion detection research community.  

When applying this approach, datasets with associated label information is crucial to not only identify malicious packets and connections, but also to construct learning models. 
However, there is a lack of relevant, publicly available datasets for researchers in the network detection intrusion community. Thus, a substantial part of past network detection intrusion studies has only relied on a small number of public data sets such as KDD Cup 1999 Data\footnote{http://kdd.ics.uci.edu/databases/kddcup99/kddcup99.html}~\cite{ahmed2016survey}.
The KDD Cup 1999 data was constructed by experts with substantial domain knowledge, 
to provide necessary connection and attack information (i.e., 41 attributes and the associated label for each connection). 
While this dataset is still relevant to this day in measuring initial performances of learning-based detection models, it is arguable that the KDD Cup 1999 data is outdated and ineffective in representing today's network, especially considering recent technological developments with a wide variety of services and applications. 


A critical part in creating datasets is the construction of labels.
Like the KDD Cup 1999 data, label construction is often done by human experts, and is thus highly laborious and expensive.
Because of this, only few public datasets have been published despite the growing importance of intrusion detection research.
In this paper, we introduce our method to construct network flow data that would aid the development of learning-based intrusion detection methods, especially concerning that in a NetFlow-compatible format. 
Our method combines the packet meta-information with IDS logs to infer labels containing intrusion information for individual  flows.
In this study, we analyze MAWILab traces that provides IDS logs with the packet meta-data~\cite{MAWILab} to generate labeled flow data.

The organization of the presentation is as follows.
In Section~\ref{sec:public_datasets}, we introduce three public datasets containing label information for network intrusion detection research.
We then transition into a two-step process which generates flow data and associated labels from the packet trace and IDS log file in Section~\ref{sec:method}. 
Our summary thus concludes our presentation in Section~\ref{sec:conc}. 

\section{Public Datasets for Network Intrusion Detection}
	\label{sec:public_datasets}

In this section, we provide a short summary of recently collected public datasets, including the KDD Cup 1999 dataset. 

\paragraph{{\bf KDD Cup 1999 dataset}}
The KDD Cup 1999 dataset contains 9-week TCP dump data collected from a local area network in 1998.
Each connection record contains the basic features of TCP connection, such as login failure, root access attempt, and others, as well as traffic features including connection error rates.
In total, there are 41 attributes from the three types of feature sets.
Along with these features, an associated label is provided that can be classified in five categories: normal, DOS (denial-of-service), R2L (unauthorized access from a remote machine), U2R (unauthorized access to local root privileges), and probing (surveillance and other probing). 
The KDD Cup 1999 dataset contains a significant number of repeated connections, which may cause biased results when evaluating intrusion detection methods. 
A modified version of this dataset, known as NSL-KDD~\cite{NSLKDD}, reduces such repeated connections to improve the data's quality. 
 
\paragraph{{\bf UNSW-NB15 dataset}~\cite{moustafa2015unsw}}
The UNSW-NB15 dataset is a more recent dataset, as it was collected in 2015. 
A traffic generator tool is configured on three servers; two of which are for the normal spread of traffic and one of which for malicious traffic.
The routers in this configuration capture packets and create pcap files.
The Bro-IDS tool\footnote{https://www.bro.org/} is utilized to generate log files from captured pcap files, along with 49 features.
The total number of records is over two millions in four CSV files. This dataset also offers a training set and testing set for evaluation purpose.

\paragraph{{\bf IDS 2017 dataset}~\cite{sharafaldin2018toward}}
The testbed to collect data includes two networks: one for attacks and the other for victims.
The captured dataset contains several different types of attacks including DoS, Web attack, infiltration attack, botnet attack, portscan, as well as background normal traffic generated by using abstract behavior of 25 users based on HTTP/HTTPS, FTP, SSH, and email protocols.
The captured data is five days long, from Monday, July 3, 2017 to Friday July 7, 2017.
The first day contains normal traffic only, while the other days have malicious activities.
The number of records is over 2.8 millions with 85 features including label information.

As seen above, all the datasets were constructed in simulated environments with dedicated networks and servers, an expensive and grueling task.
In this work, we will demonstrate how to construct datasets beneficial for intrusion detection research from the packet meta-data and IDS logs, without configuring expensive simulated environments.

\section{Data generation from MAWILab traces}
	\label{sec:method}

In this section, we introduce our method to generate flow data with associated labels for network intrusion detection research.
The generation process consists of the following two steps: 

\begin{enumerate}
	\item The first step extracts flow  information from the packet trace file (pcap); 
	\item The second step combines the IDS log data with the flow data constructed in the first step using the four-tuple of flow information (source/destination IP addresses and port numbers).
\end{enumerate}

We first describe the overview of the MAWILab data, and then present the data generation process in detail.

\subsection{Description of MAWILab data}

MAWILab provides a collection of network traffic traces and IDS logs, captured from a backbone link in Japan for about two decades up to now~\cite{mazel2014visual}\cite{callegari2016statistical}. 
The captured traces contain the TCP/IP packet header information without payloads in pcap files.
One pcap file contains 15-minute data on a single day.
The traces are publicly available to access.\footnote{http://mawi.wide.ad.jp/mawi/} 

In addition to the packet traces, it provides the  IDS logs~\cite{MAWILab}.
The log contains the label information as well as IP addresses and transport port numbers.
The labels then are inferred using a  graph-based method that compares and combines different and independent detection entities.
Table~\ref{tab:mawilab_log} shows the columns in the IDS log.
The label has three classes: 
{\em anomalous}  is for a true anomaly,
{\em suspicious}  indicates the traffic is highly likely to be anomalous,
and {\em notice}  is assigned if some detectors reported the traffic as an anomaly but it does not reach a consensus by the entire detectors.
In the current design, we only consider {\em anomalous} and {\em suspicious} in our method.

\begin{table}[!tb]
\small
\caption{MAWILab IDS log columns}
\label{tab:mawilab_log}
\centering
\begin{tabular}{ll}
\hline
Column & Description \\
\hline
sip & Source IP address \\
dip & Destination IP address \\
sport & source port \\
dport & destination port \\
taxonomy & Category of anomalies (e.g., Port scan, DoS, etc) \\
heuristic & Code assigned to  anomalies \\
	  & using  the internal heuristic \\
distance & $D_n-D_a$, \\
	 & $D_n$=distance to normal traffic, \\
	 & $D_a$=distance to anomalous traffic \\
nbDetectors & Number of detectors reported this anomaly \\
label & \{anomalous, suspicious, notice\} \\
\hline
\end{tabular}
\end{table}

\subsection{First step: generating flow data}

We construct network flow data from MAWILab packet traces using SiLK\footnote{https://tools.netsa.cert.org/silk/}.
To briefly summarize, SiLK (System for Internet-Level Knowledge) is a collection of traffic analysis tools developed to facilitate network traffic analysis.
Using this tool, it is possible to extract the flow information 
from TCP dump files in question.
Note that the output in this step will be combined with the IDS log data to generate the label information in the second step.

To obtain the flow information, 
we use the following two commands in SiLK:
{\tt rwptoflow} generates flow records from the given packet data, and
{\tt rwcut} displays the  selected fields from the SiLK flow records.
The following shows how to extract the flow information from the packet trace using {\tt rwptoflow} and {\tt rwcut}. 

{\tt \small
\begin{itemize}
	\item[$>$] rwptoflow  path  --flow-out=filename.rw 
	\item[$>$] rwcut path --fields=fieldsList  \\
             --output-path=filename.data 
\end{itemize}
}

Here, ``path'' is the directory that the output file is stored,
and ``filename'' is the output file name.
For the {\tt --fields} option, ``fieldsList'' specifies a list of the fields saved in the output file.
The list of the entire fields available 
can be found from: \url{https://tools.netsa.cert.org/silk/rwcut.html}.
Note that 
{\tt rwstats} can be alternatively used instead of {\tt rwcut}, which makes a summary output and reorders the entries (in contrast, {\tt rwcut} preserves the output order).

Here is an example. 
We downloaded a pcap file of ``201807011400.pcap.gz'' (1426.45 MB) from the trace repository.\footnote{\url{http://mawi.wide.ad.jp/mawi/samplepoint-F/2018/201807011400.html}}
Then, the following commands are executed to obtain the flow data from the given trace.
In this example, the resulted flow data is written to a file named ``20180701\_result.data''.

{\tt \small
\begin{itemize}
	\item[$>$] rwptoflow 201807011400.pcap --flow-out=20180701.rw
	\item[$>$] rwcut 20180701.rw \\
         --fields=1,2,3,4,5,6,7,8,9,10, \\
	   11,12,13,14,15,20,21,25,26,27,28,29 \\
         --output-path=20180701\_result.data
\end{itemize}
}


\subsection{Second step: combining flow data with IDS logs}

\begin{table*}[!tb]
\small
\caption{Flow features resulted from the combining process}
\label{tab:features}
\centering
\begin{tabular}{lll}
\hline
Feature	& NetFlow v9 field	& Description \\
\hline
sIP	& IPV4\_SRC\_ADDR	& Source IP address \\
dIP	& IPV4\_DST\_ADDR	& Dest IP address \\
sPort	& L4\_SRC\_PORT	& Source port \\
dPort	& L4\_DST\_PORT	& Dest port \\
proto	& PROTOCOL	& IP protocol \\
packets	& IN\_BYTES	& Packet count \\
bytes	& IN\_PKTS	& Byte count \\
flags	& TCP\_FLAGS	& Bit-wise or of TCP flags over all packets \\
sTime	& UNIX\_Seconds	& Starting time of flow (in sec) \\
durat	& 	& Duration of flow (in sec) \\
eTime	& 	& End time of flow (in sec) \\
sen	& FLOW\_SAMPLER\_ID	& Name or ID of the sensor \\
in	& SRC\_VLAN	& Router SNMP input interface \\
out	& DST\_VLAN	& Router SNMP output interface \\
nhIP	& IPV4\_NEXT\_HOP	& Router next hop ID \\
senClass		& & Class of sensor that collected flow (SiLK-specific) \\
typeFlow		& & Type of flow for this sensor class (SiLK-specific) \\
iType	& ICMP\_TYPE	& ICMP type value for ICMP flows \\
iCode	& 	& ICMP code value \\
initialF& 		& TCP flags on first packet in flow \\
sessionF& 		& Bit-wise OR of TCP flags over all packets except the first in the flow \\
attribut& 		& Flow attributes set by the flow generator \\
appli	& 	& Guess as to the content of the flow \\
class	& 	& \{normal, anomaly, unsure\} for anomaly detection \\
taxonomy & & Category of anomalies (e.g., Port scan, DoS, etc) \\
label & & \{normal, anomalous, suspicious, notice\} (MAWILab-specific)\\
heuristic & & Code assigned to  anomalies (MAWILab-specific) \\
distance & & $D_n-D_a$ (MAWILab-specific) \\
nbDetectors & & Number of detectors reported this anomaly (MAWILab-specific)\\
\hline
\end{tabular}
\end{table*}
To generate labeled flow data, the next step is to merge the output resulted from the first step with the associated IDS logs (i.e., the same-day logs as the traffic trace used in step 1).
MAWILab provides two log files for a single day traffic trace: one with a suffix of `anomalous\_suspicious' containing the log records with the labels of anomalous and suspicious, and the other with a suffix of `notice' keeping the records with the label of notice.
Again, we do not consider the label of notice since only a few detectors assumed the activity as an anomaly without making a full consent by the entire set of detectors.

To combine, we utilize the four flow attributes of \{source IP address, source port, destination IP address, destination port\}, which are available in both of the flow data file and the IDS log file. 
For each entry in the log file, we search the flow records with the identical values for the above attributes in the flow data file. 
A flow record is labeled as {\em anomaly} in case of matching.


It is possible that a log entry contains one or more null values for certain flow  attributes.
Here are two example log entries: R1:{\tt (sip=A, sport=B, dip=C, dport=D)} and R2:{\tt (sip=A, sport=null, dip=C, dport=null)}.
In this example, R2 contains null values for {\tt sport} and {\tt dport}, while R1 specifies the
entire flow attributes without a null.
For the exposition purpose, we define $L$ as the number of flow attributes available (i.e., not null) in the log entry. 
Intuitively, the log entry with a higher $L$ value is more specific.
Due to this reason, we give a higher precedence to a log entry with a higher $L$.
Thus, R1 has a greater precedence than R1; that is, R1 $>$ R2 as $L(R1) > L(R2)$.
Suppose a flow F1:{\tt (sip=A, sport=B, dip=C, dport=D)}.
It shows an example of multiple match since F1 matches both R1 and R2.
In that case, F1 is  combined with R1 by the precedence rule.

Rarely, there can be multiple matches for a single flow by multiple log entries with the same $L$.
To see this, suppose F2:{\tt (sip=P, sport=Q, dip=R, dport=S)}, R3:{\tt (sip=P, dip=R)}, and R4:{\tt (dip=R, dport=S)}. 
In this case, F2 matches both of R3 and R4 with $L$=2.
To take this into account, we apply a simple heuristic: (1) give a higher weight to victim than source (i.e., destination $>$ source, and (2) give a higher weight to host than service (i.e., IP address $>$ port number), and hence ({\tt dip $>$ sip $>$ dport $>$ sport}) for the identical $L$.
By this rule, F2 is combined with R3 instead of R4.

From IDS logs, we observed several IDS entries with $L$=1; that is, only a single attribute is available and the other three are null.
For example, one entry in the IDS log contains {\tt (sip=null, sport=443, dip=null, dport=null)}.
We also observed that 
there are about 16 million flows only matching with this record out of 68 million flows on that day (i.e., 23.5\%).
The port number 443 is widely used for secure web browser communication, and response packets from web servers communicating over TLS/SSL often have this port number.
We feel that it is somewhat risky to label such a large number of flows as {\em anomaly}.
Due to this reason, we label the flows matching more than one attribute (i.e., $L>1$) as {\em anomaly}, while we label the flows matching only one attribute ($L=1$) as {\em unsure}.
Any flow with no match is labeled as {\em normal}.
The flow records labeled as {\em unsure} can be excluded by the users based on their discretion.

Table~\ref{tab:features} shows the output format after combining.
The feature of ``class'' is the new label created in the combining process, while the feature of ``label'' shows the MAWILab-defined label information. 
The table also shows how the features in the output are associated with the fields defined in NetFlow v9.

We implemented this combining process using Python.
A Python program {\tt flowlabeling.py} takes a flow data file (resulted in step 1) and an IDS log file, and produces a set of combined flows formatted in Table~\ref{tab:features}.
Another Python program {\tt flowsplitter.py} breaks the outputs into multiple files with designated time windows.
For example, it splits a 15-minute flow data into 180 sub-files under the assumption of 5-second time window. 
The programs are available from the following repository: \url{https://github.com/dcstamuc/FlowDataGen}.

\section{Summary} 
	\label{sec:conc}	

This paper introduced a method to construct network flow data that would aid the development of learning-based intrusion detection methods.
Our method combines the packet meta-information with the IDS logs to infer labels containing intrusion information for individual network flows.
To achieve this goal, we utilized the SiLK tool to extract the flow data from the TCP dump file, and implemented a Python program to combine the flow data with the IDS log.
The generated flow data contains associated label information for intrusion detection research and is NetFlow compatible.
We believe the introduced method would assist researchers in network intrusion detection to access recent network flow datasets with associated labels.

%

\bibliographystyle{unsrt}
\bibliography{paper}

\end{document}